\documentclass[12pt]{article}

\usepackage{endnotes}
\let\footnote=\endnote
\usepackage{graphicx,amssymb,amsmath,amsthm,cite,cases}
\usepackage{verbatim}
\DeclareGraphicsExtensions{.jpg}
\usepackage{algorithmic}
\usepackage[ruled]{algorithm}  

\def\expn{e^{\imath \frac{\pi(n-1)}{M}}}
\def\expin{e^{\imath \frac{2\pi(j-1)(n-1)}{M}}}
\def\cexpin{e^{-\imath \frac{2\pi(j-1)(n-1)}{M}}}
\def\expiln{e^{\imath \frac{2\pi(j-1)(\ell_n-1)}{M}}}
\def\iset{j=1,\ldots,N}

\def\be{\begin{equation}}
\def\ee{\end{equation}}
\def\ben{\begin{eqnarray}}
\def\een{\end{eqnarray}}

\def\ds{\displaystyle}

\def\D{\mathcal{D}}
\def\R{\mathbb{R}}
\def\N{\mathbb{N}}
\def\C{\mathbb{C}}
\def\V{\mathbb{V}}

\def\B{\mathcal{B}}
\def\til{\tilde}
\def\ct{\til{c}}
\def\vct{\mathbf{\ct}}
\def\ov{\overline}

\def\bel{\begin{linenomath}}
\def\eel{\end{linenomath}}

\def\vR{\mathbf{R}}
\def\vRt{\mathbf{\til{R}}}
\def\vd{\mathbf{d}}
\def\vc{\mathbf{c}}
\def\vf{\mathbf{f}}

\def\vy{\mathbf{y}}

\def\vB{\mathbf{b}}

\def\vRe{\mathbf{R}}
\def\vIP{\mathbf{IP}}

\def\Aux{\mathbf{Aux}}

\def\Re{\operatorname{Re}}
\def\Im{\operatorname{Im}}
\def\im{\operatorname{\imath}}

\def\op{\hat{P}}
\def\J{\hat{J}}
\def\kq{k_q}
\def\Ns{N_b}

\newcommand{\la}{\langle}
\newcommand{\ra}{\rangle}

\newcommand{\Spann}{{\mbox{\rm{span}}}}

\begin{document}
\title{A dedicated greedy pursuit algorithm\\ 
for sparse spectral representation of music sound}
\author{Laura Rebollo-Neira and Gagan Aggarwal\\
Mathematics Department\\
Aston University\\
B3 7ET, Birmingham, UK\\
email: \texttt{l.rebollo-neira@aston.ac.uk}}
\maketitle
\baselineskip = 1.7 \baselineskip
\newpage
\begin{center}
\textbf{Abstract}
\end{center}
A dedicated algorithm for sparse spectral representation of 
music sound is presented.
The goal is to enable the representation of a 
piece of music signal as a linear superposition 
of as few spectral components as possible, without 
affecting the quality of the reproduction. 
A representation of this nature 
is said to be sparse. In the present context 
sparsity is accomplished by greedy selection of the spectral 
components, 
from an overcomplete set called a 
{\em{dictionary}}. The proposed algorithm is tailored to be 
applied with trigonometric dictionaries.
 Its distinctive feature being that it avoids the need for 
the actual construction of the whole dictionary,
by implementing the 
required operations via the Fast Fourier Transform. 
The achieved sparsity is theoretically equivalent to 
that rendered by the Orthogonal Matching Pursuit 
method. The contribution of the proposed dedicated 
implementation is to extend the applicability of 
 the standard Orthogonal Matching Pursuit algorithm, 
 by reducing its storage and computational demands.
The suitability of the approach for producing 
sparse spectral representation is illustrated by 
 comparison with the traditional method, in the line of 
the Short Time Fourier Transform, involving only 
 the corresponding orthonormal trigonometric basis. 
\vspace*{.3in}


\newpage
\section{Introduction}
\label{Int}
Spectral representation is a classical approach which
plays a central role in the analysis and modelling 
of both, music sounds (Serra and Smith, 1990; 
Fletcher and Rossing, 1998; Davy and Godsill, 2003)
and acoustic properties of music instruments 
(Wolfe {\em{et al.}}, 2001). 

Available techniques aiding the 
spectral analysis of music range from the 
Fast Fourier Transform (FFT) 
and Short Time Fourier Transform (STFT) 
to several classes of 
joint Time Frequency$/$Scale distributions (Alm and 
Walker, 2002; Smith 2011)  and 
 atomic representations (Mallat and Zhang, 1993; 
Gribonval and Bacry, 2003).

In this Communication we focus on the representation   
of a digital piece of music, as the 
superposition of vectors arising by the discretization 
of trigonometric functions. The aim is to represent
segments of a sound signal, as a linear combination 
of as few spectral components as possible without 
affecting the quality of the sound reproduction.
We referrer to the sought representation as 
{\em{piecewise sparse spectral representation}} 
of music sound. 
Additionally to the typical advantages of sparse
signal representation, the emerging 
theory of compressive$/$compressed sensing
(Baraniuk, 2007, 2011;  Donoho, 2006; 
 Cand\`{e}s, {\em{et al.}} 2006; 
Cand\`{e}s and Wakin, 2008)
 has introduced a renewed strong reason to pursue 
sparse  representation of music. This theory associates 
sparsity to a new framework for digitalization,
beyond the Nyquist$/$Shannon sampling theorem. Within 
the compressive sensing framework, the number of 
measurements needed for accurate representation of 
a signal informational content 
decreases, if the sparsity of the representation improves.

For the class of compressible signals the 
 sparse approximation  
can be accomplished by 
representation in an orthonormal basis, simply 
by disregarding 
the least significant terms in the decomposition. 
Melodic music signals 
are known to be compressible in terms of 
trigonometric orthonormal basis.  However, a much higher
level of sparsity may be achieved by releasing 
the orthogonality property of the spectral components
(Mallat and Zhang, 1993; Gribonval and Bacry, 
2003; Rebollo-Neira, 2016a).
The price to be paid for that is the increment in the  
complexity of the numerical 
algorithms producing the corresponding 
sparser approximation. Practical algorithms for  
this purpose are known as greedy pursuit strategies
(Friedman and Stuetzle, 1981; Jones, 1987;
 Mallat and Zhang, 1993). 
In Gribonval and Bacry (2003) a dedicated Matching Pursuit 
method for effective implementation of 
the spectral model is developed by means of well localized 
frequency components of variable length. 
In Rebollo-Neira (2016a) an alternative approach is 
considered. It involves the approximation of a signal by 
partitioning, according to the following steps: 
i)The signal is divided into 
small units (blocks) ii)Each block is approximated 
 by nonorthogonal spectral components, independently of 
each other but somewhat `linked' by a 
global constraint on sparsity or quality. The 
global constraint is 
 fulfilled by establishing a hierarchy for the order 
in which each element in the partition 
is to be approximated. Thus, the method requires 
significant storage. Even if 
the global constraint is disregarded, 
and each unit approximated totally independent of 
the others, the algorithms in Rebollo-Neira (2016a) are 
effective for partition units of moderate length. 
For units of larger size there is a need of mathematics 
algorithms specialized to that situation. 
This is the goal of the 
present work. We propose a dedicated algorithm for 
 nonorthogonal sparse spectral modeling which, 
as a consequence of allowing for relatively 
large elements in a partition, somewhat reduces the 
need for a 
global constraint on sparsity. This makes it possible  
for the 
approximation of each unit up to the same quality and 
completely independent of the others. 
The approach is, thereby, suitable 
for straightforward parallelization in multiprocessors.  
As far as sparsity is concerned, the results are theoretical 
equivalents to those produced by the effective 
Orthogonal Matching Pursuit method (Pati {\em {et al.}}, 
1993). The particularity of the proposed implementation, 
dedicated to trigonometric dictionaries, is that 
it avoids the need for storing the whole dictionary and 
reduces the complexity of calculations via the  
 Fast Fourier Transform. The relevance of sparse 
spectral representation with 
trigonometric dictionaries, 
in the context of music compression 
 with high quality recovery, is illustrated 
in Rebollo-Neira (2016b).

The paper is organized as follows: 
Sec.~\ref{SSM} discusses the spectral model outside 
the traditional orthogonal framework. The  
 mathematical methods for operating within the 
nonorthogonal setting are also discussed in this section,
motivating the proposed  dedicated 
approach. The approach is first explained and 
then summarized in the form of pseudocodes (Algorithms 
1-6) given in Appendix A. The examples of Sec.~\ref{NE} 
illustrate the benefit of a nonorthogonal framework,
against the orthogonal one, 
in relation to the very significant gain in the 
sparsity of the spectral representation of music signals 
for high quality recovery.
The results presented in this section demonstrate the 
relevance of the proposed greedy strategy dedicated to  be 
applied with trigonometric dictionaries.
The conclusions are summarized in Sec.~\ref{Con}.  
\section{Sparse Spectral Representation} 
\label{SSM}
Let's assume that a sound signal is given 
by $N$ sample values, $f(j),\,j=1,\ldots,N$, which are
modeled by the following transformation:
\be
f(j)= \frac{1}{\sqrt{N}} 
\sum_{n=1}^M c(n) \expin, \quad j=1,\ldots,N.
\label{mof}
\ee
For $M=N$ the set of vectors 
 $\{\frac{1}{\sqrt{N}}\expin,\, j=1,\ldots,N\}_{n=1}^M$ 
is an orthonormal basis for 
the subspace of $N$-dimensional vectors of
complex components. Thus the coefficients in \eqref{mof} 
are easily obtained as 
\be
c(n) =\frac{1}{\sqrt{N}} \sum_{j=1}^M f(j) \cexpin, 
\quad n=1,\ldots,M=N.
\label{cf}
\ee
Equations \eqref{mof} and \eqref{cf} can be evaluated in a 
fast manner via the FFT. 

Suppose now that $M>N$. In that case the set 
$\{\frac{1}{\sqrt{N}}\expin,\, j=1,\ldots,N\}_{n=1}^M$
 is no longer an orthonormal basis 
but a {\em{tight frame}} (Young, 1980, Daubechies, 1992). From a 
computational viewpoint the difference with the case 
$M=N$ is much less pronounced than the theoretical 
difference.
Certainly, when dealing with a tight frame the 
  coefficients in \eqref{mof} can still 
be calculated via FFT, by zero padding. The 
differences though with the orthogonal case are major.
\begin{itemize}
\item[i)]
When $M>N$ the coefficients in the superposition 
\eqref{mof} are {\underline{not}} unique. 
The addition of a linear combination with coefficients 
taken as the components of any vector in the null space 
of the transformation would not affect the 
reconstruction.
\item[ii)]
The tight frame coefficients calculated via FFT,  by
zero padding, produce the unique coefficients 
 minimizing the square norm $\ds{{\sum_{n=1}^M |c(n)|^2}}$.
Such a solution 
is {\underline{not}} sparse. 
\item[iii)]
For the case $M=N$ the approximation obtained through 
\eqref{mof}, by disregarding 
coefficients of small magnitude, is optimal in the 
sense of minimizing the norm of the residual error. 
This is {\underline{not}} true when $M>N$, in  which 
case the nonzero 
coefficients need to be re-calculated to attain
the equivalent optimality (Rebollo-Neira, 2007).
\end{itemize}
In order to construct an optimal approximation of the  data 
 by a representation of the form 
 \eqref{mof}, with  $M>N$ 
 but containing at most $k$ non zero coefficients, 
those coefficients have to be appropriately calculated.
Let's suppose that we want to involve only the elements 
$\ell_n,\,n=1,\ldots,k$ where each $\ell_n$ is a different 
member from the set $\{1,2,\cdots,M\}$.  Then the 
approximation model takes the form
\be
\label{amo}
f^k(j)=\frac{1}{\sqrt{N}} \sum_{n=1}^k c^k(\ell_n) \expiln,\quad \iset.
\ee
The superscript $k$ in the coefficients 
$c^k(\ell_n),\,n=1,\ldots,k$ indicates that they have 
to be recalculated if 
some terms are added to (or eliminated from) the
model  \eqref{amo}.
We address the matter of 
 choosing the $k$ elements in \eqref{amo}
by a dedicated  
Self Projected Matching Pursuit (SPMP) approach 
(Rebollo-Neira and Bowley, 2013).
\normalfont{\subsection{{Self Projected Matching Pursuit}}}
Before reviewing the general SPMP technique let's 
define some basic notation: $\R, \C$ and $\N$
represent the sets of real, complex 
and natural numbers, respectively. 
Boldface letters are used to indicate Euclidean vectors
 and standard mathematical fonts for their
components, e.g., $\vd \in \C^N$ is a vector of 
$N$-components
$d(j) \in \C^N\,, j=1,\ldots,N$. The operation 
$\la \cdot,\cdot \ra$ indicates the Euclidean inner
product and  $\| \cdot \|$ the induced norm, i.e. 
$\| \vd \|^2= \la \vd, \vd \ra$, with the usual 
inner product definition: For $\vd \in \C^N$ 
and $\vf \in \C^N$
$$ \la \vf, \vd \ra = \sum_{j=1}^N f^\ast(j) d(j),$$
where $f^\ast(j)$ stands for the complex conjugate of 
$f(j)$.

Let's consider now a set $\D$ of $M$ normalized 
to unity vectors 
$\D=\{\vd_n \in \C^{N}\,; \|\vd_n\|=1\}_{n=1}^M$ 
spanning $\C^N$. For $M>N$    
 the over-complete set $\D$ is called a dictionary 
 and the elements are called {\em{atoms}}. 
Given a signal, as a vector $\vf \in \C^{N}$, the 
$k$-term {\em{atomic decomposition}} for its
approximation  takes the form
\be
\label{atoq}
\vf^{k}= \sum_{n=1}^{k}
c^{k}(\ell_n) \vd_{\ell_n}.
\ee
The problem of how to select from $\D$ the 
$k$ elements $\vd_{\ell_n},\,n=1\ldots,k$, such that 
$\|\vf^k - \vf\|$ is minimal, is an NP-hard  problem
(Natarajan, 1995). The equivalent problem,  that of 
 finding the 
sparsest representation for a given upper bound error, is
also NP hard. Hence, in practical applications 
 one looks for `tractable sparse' solutions.
This is 
a representation  involving a number of $k$-terms, 
 with  $k$  acceptable small in relation to $N$. Effective
 techniques 
available for the purpose 
 are in the line of Matching Pursuit
Strategies. The seminal approach, Matching Pursuit (MP), 
was introduced with this name in the context of
signal processing by Mallat and Zhang (1993).
Nevertheless, it had appeared previously 
as a regression technique in 
statistics (Friedman and Stuetzle, 1981)
 where the convergence property
was established (Jones, 1987). 
The MP implementation  is
very simple. It 
evolves by successive approximations as follows.

Let $\vRe^{k}$ be the $k$-th order residue defined as
$\vRe^{k}= \vf -\vf^k$, and $\ell_{k+1}$ the index for
which the corresponding dictionary atom $\vd_{\ell_{k+1}}$ yields a maximal value of $|\la \vd_{n} , \vRe^{k} \ra|,\, n = 1,\ldots M$.
Starting with an initial approximation
$\vf^0=0$ and $\vRe^{0} = \vf - \vf^0$
the algorithm
iterates by sub-decomposing the $k$-th order residue into
\be
\vRe^{k} =
\la \vd_{\ell_{k+1}}, \vRe^{k} \ra \vd_{\ell_{k+1}} + \vRe^{k+1}, 
\label{tech:1}
\ee
which defines the residue at order $k+1$.
Because the atoms are normalized to unity 
$\vRe^{k+1}$ given in \eqref{tech:1} is orthogonal to 
$\vd_{\ell_{k+1}}$.  Hence it is true that
\be
\|\vRe^{k}\|^{2} = |\la \vd_{\ell_{k+1}}, \vRe^{k} \ra|^{2} + \|\vRe^{k+1}\|^{2},
\quad  n = 1,\ldots, M,
\label{tech:2}
\ee
from where one gathers that the 
 dictionary atom $\vd_{\ell_{k+1}}$ yielding a maximal value
of $|\la \vRe^{k}, \vd_{n} \ra|$ minimizes $\| \vRe^{k+1}\|^{2}$. Moreover, it  follows  from \eqref{tech:1} that 
at iteration $k$ the MP
algorithm results in an intermediate representation of
 the form:
\be
\vf= \vf^k + \vRe^{k+1},
\label{tech:3}
\ee
with
\be
\vf^{k}= \sum_{n=1}^{k} \la \vd_{\ell_{n}}, \vRe^{n-1} \ra \vd_{\ell_{n}}.
\label{tech:4}
\ee
In the limit $k \rightarrow \infty$ the sequence  $\vf^{k}$
converges to $\vf$, or to  $\hat{P}_{\V_{M}}\vf $,
the orthogonal projection of $\vf$ onto
$\V_{M}= \Spann\{\vd_{\ell_{n}}\}_{n=1}^M$ if $\vf$ 
were not in $\V_{M}$ (Jones, 1987;
Mallat and Zhang, 1993; Partington 1997).  
Nevertheless, if the algorithm is stopped at the 
$k$th-iteration,
$\vf^{k}$ recovers an approximation of $\vf$ with an
error equal to the norm of the residual $\vRe^{k+1}$ which, 
if the selected atoms are not
orthogonal, will not be orthogonal to the subspace
 they span. An additional drawback of the MP approach 
is that the selected atoms may not be linearly independent.
As illustrated in Rebollo-Neira and Bowley (2013), 
this drawback may significantly compromise sparsity
 in some cases.
A refinement to MP, which does yield an
orthogonal projection approximation at each step, has been termed
Orthogonal Matching Pursuit (OMP) (Pati {\em {et al.}}, 1993).
In addition to selecting only linearly independent atoms,
the OMP approach improves upon MP numerical convergence rate
and therefore amounts to be, usually,
a better approximation of a signal after a finite number
 of iterations.
OMP provides a decomposition of the signal of the form:
\begin{equation}
\vf = \sum_{n=1}^{k} c^k(\ell_n) \vd_{\ell_{n}} + \tilde{\vRe}^{k},
\label{tech:5}
\end{equation}
where the coefficients $c^k(\ell_n)$ are computed 
 to guarantee that 
\begin{equation}
\sum_{n=1}^{k} c^k(\ell_n) \vd_{\ell_{n}}= 
\hat{P}_{\V_{k}}\vf,\quad{\text{with}}
\quad \V_{k}= \Spann\{\vd_{\ell_{n}}\}_{n=1}^k.
\label{e9}
\end{equation}
The coefficients giving rise to the 
orthogonal projection $\hat{P}_{\V_{k}}\vf$
 can be calculated as 
$c^k(\ell_n) =\la \vB_n^{k}, \vf \ra$, 
where the vectors $\vB_{n}^{k},\,n=1,\ldots,k$ 
 are biorthogonal to the selected atoms 
$\vd_{\ell_n},\, ,n=1,\ldots,k$ and span
 the identical subspace, i.e.,
${\ds{\V_{k}= \Spann\{\vB_{n}^{k}\}_{n=1}^{k}=
 \Spann\{\vd_{\ell_n}\}_{n=1}^{k}}}$.
These coefficients yield the unique element 
$\vf^k \in \V_{k}$ minimizing
$\|\vf^k -\vf\|$.   
A further optimization of MP, called
Optimized Orthogonal Matching Pursuit (OOMP) 
 improves on OMP by also selecting the atoms 
 yielding stepwise minimization  
 of $\|\vf^k -\vf\|$ (Rebollo-Neira and Lowe, 2002). 
Both OMP and OOMP are very effective
approaches for processing signals up to
some dimensionality.
They become inapplicable, due to its storage 
requirements, when the signal 
dimension exceeds some value.
Since large signals are approximated by partitioning, 
 up to some size of the partition unit both OMP and 
 OOMP are suitable tools. For considering units of 
 size exceeding the limit of OMP applicability, the 
 alternative implementation, SPMP, 
 which yields equivalent results 
(Rebollo-Neira and Bowley, 2013) is to be applied.
The latter is based on the fact that, as already mentioned,
the seminal MP approach converges asymptotically
to the orthogonal projection onto
the span of the selected atoms. 
Hence MP itself can be used to produce an
orthogonal projection of the data, at each iteration, 
by self-projections. The orthogonal projection  is
realized by subtracting from the residue its approximation 
constructed through the MP approach, 
but only using  the already selected atoms as dictionary. 
 This avoids the need of computing and storing the 
above introduced vectors $\vB_{n}^{k},\,n=1,\ldots,k$, 
for calculating the coefficients in \eqref{e9}. 

The SPMP method progresses as follows (Rebollo-Neira and Bowler, 2013).
Given a dictionary 
$\D=\{\vd_n \in \C^{N}\,; \|\vd_n\|=1\}_{n=1}^M$
  and a signal $\vf  \in \C^{N}$,
set $S_0=\{\emptyset\}$, $\vf^0=0$, and $\vR^0=\vf$. Starting with  
 $k=1$, at each iteration $k$ implement the steps below.
\begin{itemize}
\item[i)]
Apply the MP criterion described above for selecting 
one atom from $\D$, i.e., select $\ell_{k}$ such that 
\be
\label{selMP}
\ell_{k}=  \operatorname*{arg\,max}_{\substack{n=1,\ldots,M}} |\la \vd_{n} , \vRe^{k-1} \ra|
\ee
and assign $S_k = S_{k-1} \cup \vd_{\ell_{k}}$. 
Update the approximation of $\vf$ as 
 $\vf^k = \vf^{k-1} + 
\la \vd_{ \ell_{k}} , \vRe^{k-1} \ra \vd_{ \ell_{k}}$ 
and evaluate the new residue  $\vRe^k= \vf -\vf^k$.
\item[ii)]
Approximate $\vRe^k$ using only the selected set
$S_k$ as the dictionary, which guarantees the asymptotic
convergence to the
approximation $\op_{\V_k}{\vRe^k}$ of $\vRe^k$, 
where $\V_k= \Spann \{ S_k\}$,
and a residue $\vRe^\perp= {\vRe^k} - \op_{\V_k}\vRe^k$ 
having no component in $\V_k$.
\item[iii)]Set $\vf^k \leftarrow  \vf^k + \op_{\V_k}\vRe^k, 
 \vRe^k \leftarrow \vRe^\perp,  k \leftarrow k+1$, 
 and repeat steps
i) - iii) until, for a required  $\rho$, the
condition $\|\vRe^k\| < \rho$ is reached.
\end{itemize}
\subsection{Dedicated SPMP algorithm for 
sparse spectral decomposition}
\label{Sec12}
Even if SPMP reduces the storage requirements for 
 calculating and adapting the coefficients of an atomic 
decomposition, storage and 
complexity 
remains an issue for processing a signal by partitioning in 
units of considerable size. 
Notice that the SPMP method involves repetitive calculations of inner 
products.
The advantage of using a trigonometric dictionary,
 in addition 
to rendering highly sparse representations in relation to 
a trigonometric basis, is that a trigonometric dictionary
allows the design of a dedicate SPMP implementation, 
which avoids the construction 
and storage of the actual dictionary by calculating 
inner products via FFT.

From now on we shall make use of the knowledge that a 
piece of music is given by real numbers, 
i.e. $\vf \in \R^N$. 
The dictionaries 
we consider for producing sparse spectral decompositions 
of the data are: 
the Redundant Discrete Fourier (RDF) dictionary, $\D^f$, 
the Redundant Discrete Cosine (RDC) dictionary, $\D^c$,
and the Redundant Discrete Sine (RDS) dictionary, $\D^s,$
defined below.
\begin{itemize}
\item
$\mathcal{D}^{f}=\{\frac{1}{\sqrt{N}}\,
 e^{\imath   \frac{{2 \pi(j-1)(n-1)}}{M}},\, j=1,\ldots,N\}_{n=1}^{M}.$
\item
$\mathcal{D}^{c}=\{ \frac{1}{w^c(n)}\,
 \cos ({\frac{{\pi(2j-1)(n-1)}}{2M}}),\,j=1,\ldots,N\}_{n=1}^{M}.$
\item
$
\mathcal{D}^{s}=\{\frac{1}{w^s(n)}\, \sin  ({\frac{{\pi(2j-1)n}}{2M}}), \,j=1,\ldots,N\}_{n=1}^{M},$
\end{itemize}
where $w^c(n)$ and $w^s(n),\, n=1,\ldots,M$ are 
normalization factors as given by
$$w^c(n)=
\begin{cases}
\sqrt{N} & \mbox{if} \quad n=1,\\
\sqrt{\frac{N}{2} + \frac{ 
\sin(\frac{\pi (n-1)}{M}) \sin(\frac{2\pi(n-1)N}{M})}
{2(1 - \cos(\frac{ 2 \pi(n-1)}{M}))}} & \mbox{if} \quad n\neq 1.
\end{cases}
$$
$$w^s(n)=
\begin{cases}
\sqrt{N} & \mbox{if} \quad n=1,\\
\sqrt{\frac{N}{2} - \frac{ 
\sin(\frac{\pi n}{M}) \sin(\frac{2\pi n N}{M})}
{2(1 - \cos(\frac{ 2 \pi n}{M}))}} & \mbox{if} \quad n\neq 1.
\end{cases}
$$
For $M=N$ each of the
above dictionaries is an orthonormal basis, 
the Orthogonal Discrete Fourier (ODF), Cosine (ODC), and
Sine (ODS) basis, henceforth to be denoted  as 
$\B^f$ $\B^c$ and
$\B^s$ respectively. The joint mixed dictionary
$\D^{cs} = \D^c \cup \D^s$,  with $\D^c$ and $\D^s$
having the same number of 
elements,
is an orthonormal
basis for $M=\frac{N}{2}$, the Orthogonal Discrete 
Cosine-Sine (ODCS) basis  to be indicated as $\B^{cs}$. 
If $M> \frac{N}{2}$, 
 $\D^{cs}$ becomes a Redundant Discrete Cosine and Sine (RDCS) dictionary. 

For facilitating the discussion of fast calculation
 of inner products
with trigonometric atoms, given a vector $\vy\in \C^N$, 
let's define
\be
\label{dft}
{\cal{F}}(\vy,n,M)= \sum_{j=1}^N y(j) e^{-\im 2 \pi \frac{(n-1)(j-1)}{M}},\quad n=1,\ldots,M.
\ee
When $M=N$ \eqref{dft} is the Discrete Fourier Transform
of vector $\vy \in \C^N$, which can be evaluated using
FFT. If $M>N$ we can still calculate \eqref{dft} via
FFT by padding with $(M-N)$ zeros the vector  $\vy$.
Equation \eqref{dft} 
  can also be used to calculate inner
products with the atoms in dictionaries  $\D^c$ and
$\D^s$. Indeed,
\be
\label{dct}
\sum_{j=1}^{N}\cos{\frac{{\pi(2j-1)(n-1)}}{2M}}
y(j)= \Re \left(e^{-\im \frac{\pi (n-1)}{2M}}
{\cal{F}}(\vy,n,2M)\right),\,
n=1,\ldots,M.
\ee
and
\be
\label{dst}
\sum_{j=1}^{N}\sin{\frac{{\pi(2j-1)(n-1)}}{2M}}
y(j)= - \Im \left(e^{-\im \frac{\pi (n-1)}{2M}}
{\cal{F}}(\vy,n,2M)\right), \,
n=2,\ldots, M+1,
\ee
where $\Re(z)$ indicates the real part of $z$,    
$\Im(z)$ its imaginary part,  and the 
notation ${\cal{F}}(\vy,n,2M)$ implies that the 
vector $\vy$ is padded with $(2M-N)$ zeros.               

We associate the dictionaries                   
$\D^{f}, \D^{c}, \D^{s}$ and $\D^{cs}$              
to the cases I, II, III, and IV,                   
of the dedicated SPMP Algorithm (SPMPTrgFFT), 
which is developed in Algorithm~6 of Appendix A, by 
recourse to the procedures  given in 
Algorithms~1-5.
\subsection{Procedures for an implementation of   
the SPMP method  dedicated to trigonometric 
dictionaries}
Let us recall once again that the aim of the 
present work is to be able to apply the 
SPMP algorithm, witch is theoretically equivalent 
to the OMP method, 
but without evaluating and storing the dictionaries
$\D^{f}, \D^{c}, \D^{s}$ or $\D^{cs}$. Instead, only the 
selected atoms are evaluated (Algorithm \ref{TrgAt}) 
and the inner products are
performed via FFT (Algorithm \ref{IP}). 
Apart from that, the  dedicated implementation follows
the steps of the general SPMP method. Some particular 
features are worth remarking.
\begin{itemize}
\item
Notice that for Case I,
as a consequence of the data being real numbers,
it holds that ${\cal{F}}(\vy,n,M)= {\cal{F}}^\ast(\vy, 
M- n +2,M)$. Hence 
the atoms can be taken always
in pairs, $\ell_k$ and $(M- \ell_k +2)$.
\item
The procedure for self projection of MP 
 (Algorithm \ref{ProjMP}),
is a recursive implementation of the selection procedure, 
but the selection is carried out only over the, say $k$, 
 already selected atoms (Algorithm \ref{AlReSel}). 
Then the calculation of the relevant inner products is 
worth being carried out via FFT only for 
values of $k$ larger than
${\ds{\frac{M}{N} \log_2M}}$. 
\item
In order to provide all the implementation details
of the proposed method in a clear and testable manner,
we have made publicly available a MATLAB
version of the pseudocodes (Algorithms 1-6),
as well as the script and the signals  
which will allow the interested researcher to reproduce 
the numerical 
results in this paper \footnote{http://www.nonlinear-approx.info/examples/node02.html}.
The MATLAB routines should be taken only
 as `demonstration material'. They are not intended
to be an optimized
implementation of the algorithms. Such optimization
should depend on the programming language used for
practical applications.
\end{itemize}
\section{Numerical Examples}
\label{NE}
We apply now the SPMPTrgFFT algorithm to produce a 
sparse  spectral representation of the sound clips  
listed in Table~1 and Table~2.
The approximation is carried out by dividing the signals 
into disjoint pieces $\vf_q \in \R^{\Ns},\, 
q=1,\ldots,Q$ of uniform length $\Ns$, i.e., 
 $\vf = \J_{q=1}^{Q} \vf_q$, where $\J$ 
indicates a concatenation operation and $N=Q \Ns$.

The purpose of the numerical example is to illustrate 
 the relevance of the method to produce
sparse spectral representation of music, in comparison 
to the classical orthogonal representation in the line 
of STFT.
Each segment $q$ is approximated up to the same quality. 
The sparsity is measured by the Sparsity Ratio (SR) 
defined as
$\ds{\text{SR}= \frac{N}{K}}$, where $K$ is the total 
number of coefficients in the signal representation, i.e,
denoting by $\kq$ the number of coefficients for 
approximating the $q$-th segment $K=\sum_{q=1}^Q \kq$. 
 
As a measure of approximation
quality we use the standard Signal to Noise Ratio
(SNR),
$$\text{SNR}=10 \log_{10} \frac{\| \vf\|^2}{\|\vf - \vf^k\|^2}=
10 \log_{10}\frac{\sum_{\substack{j=1\\q=1}}^{\Ns,Q} |f_q(j)|^2}
{\sum_{\substack{j=1\\q=1}}^{\Ns,Q} |f_q(j) -f^\kq(j)|^2}.$$
All the clips of Table~1 are approximated up to  
SNR=35dB.
The approximation has been carried 
out using all the dictionaries introduced 
in Sec.~\ref{Sec12}, with redundancy four, 
and all the concomitant orthogonal basis. Due to space 
limitation only the best results produced  by a dictionary, 
and by a basis, are reported.
The best dictionary results are rendered  
by the mixed dictionary $\D^{cs}$. Nevertheless, in the 
case of a basis the best results are achieved by the 
cosine basis $\B^c$. 
\begin{figure}[!ht]
\begin{center}
\includegraphics[width=8cm]{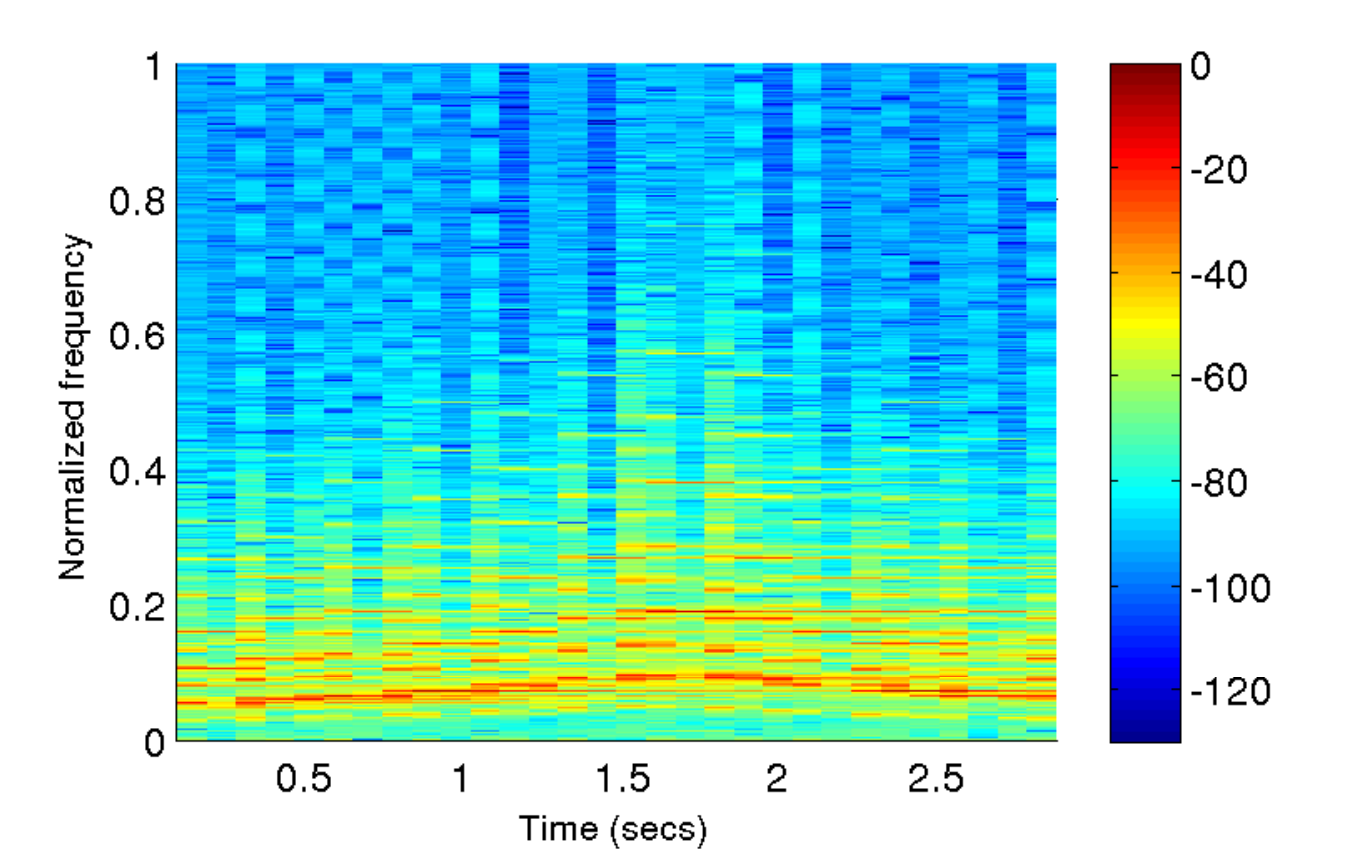}
\includegraphics[width=8cm]{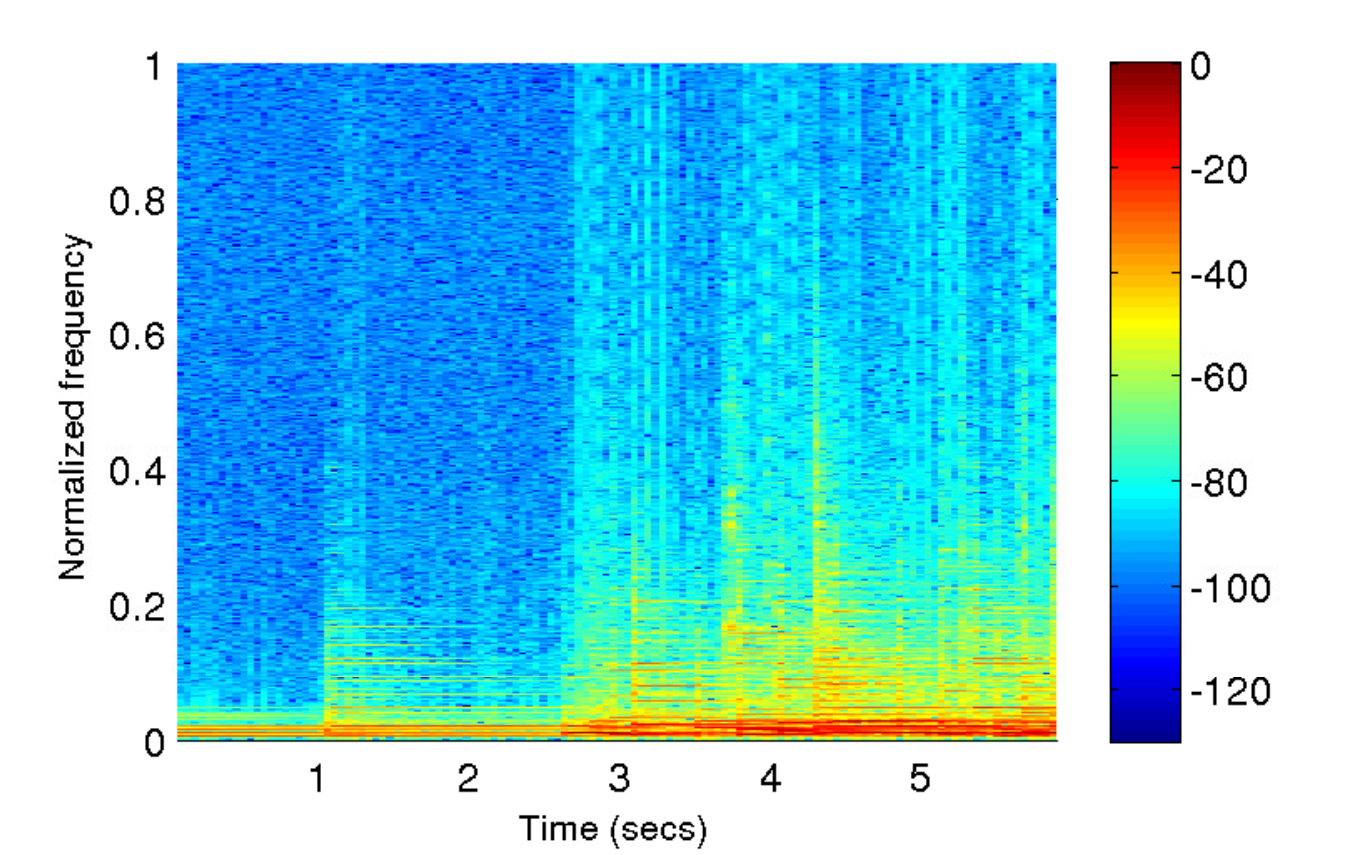}
\caption{\small{(Color online only) Spectrograms of the Flute Exercise clip (left)  
$N=65536$ samples at 22050 Hz, and that of the 
Classic Guitar, $N=262144$ samples at 44100Hz. Each 
spectrogram was produced using a Hamming window
of length $4096$ samples and 50$\%$ overlap.}}
\end{center}
\end{figure}
The approximation of all the clips in Table~1 was 
carried out for partitions corresponding to 
$\Ns$ equal to 512, 1024, 2048, 4096, 8192,
and 16384 samples. 
For space limitation 
only the sparsity results corresponding to 
all those values of $\Ns$ are shown for the first two
clips of the table.
Fig.~1 gives the classic spectrogram for the
Flute Exercise and Classical Guitar.
Fig.~2  shows the values of 
the SR for those clips,  
as a function of the partition unit size 
$\Ns$.
\begin{figure}[!ht]
\begin{center}
\includegraphics[width=8cm]{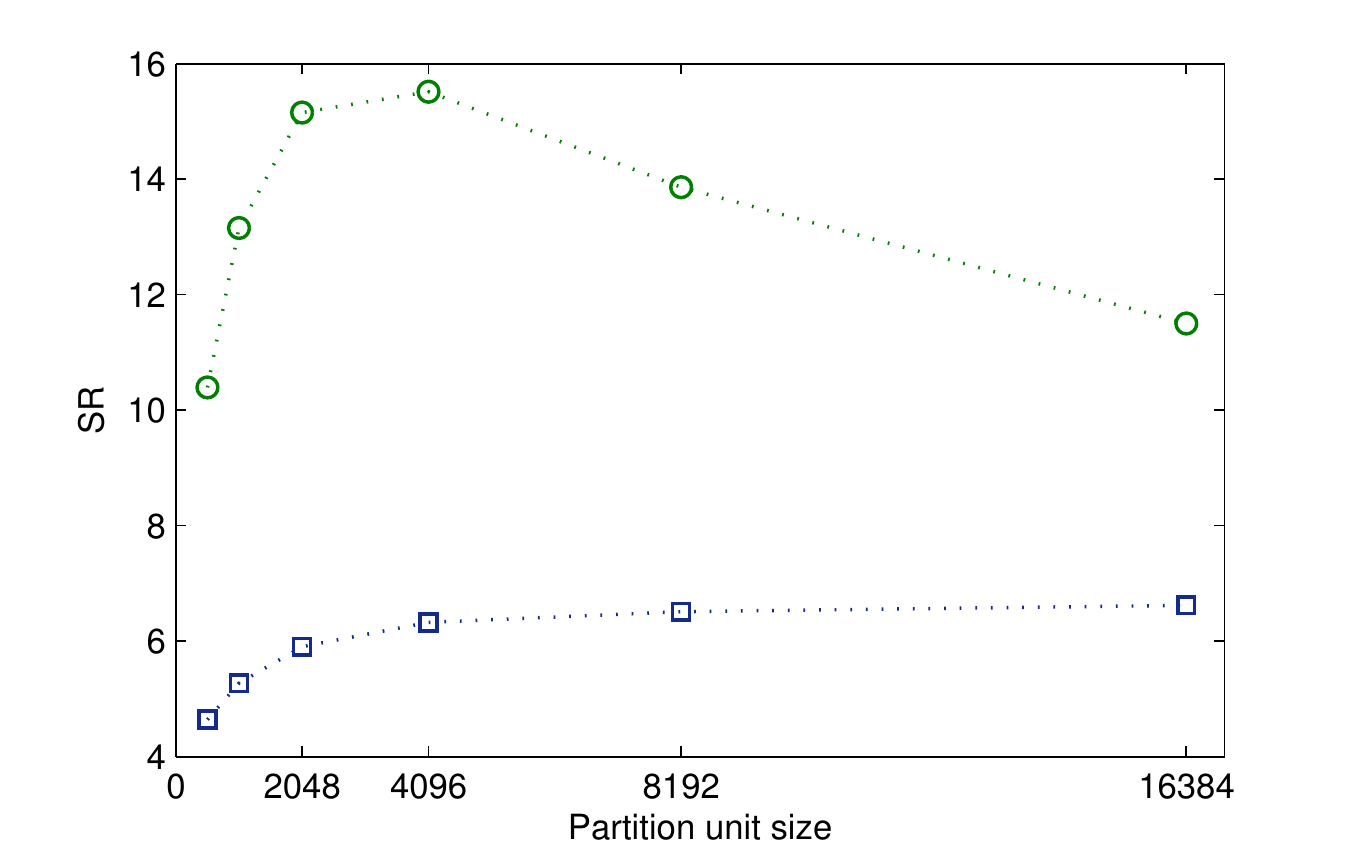}
\includegraphics[width=8cm]{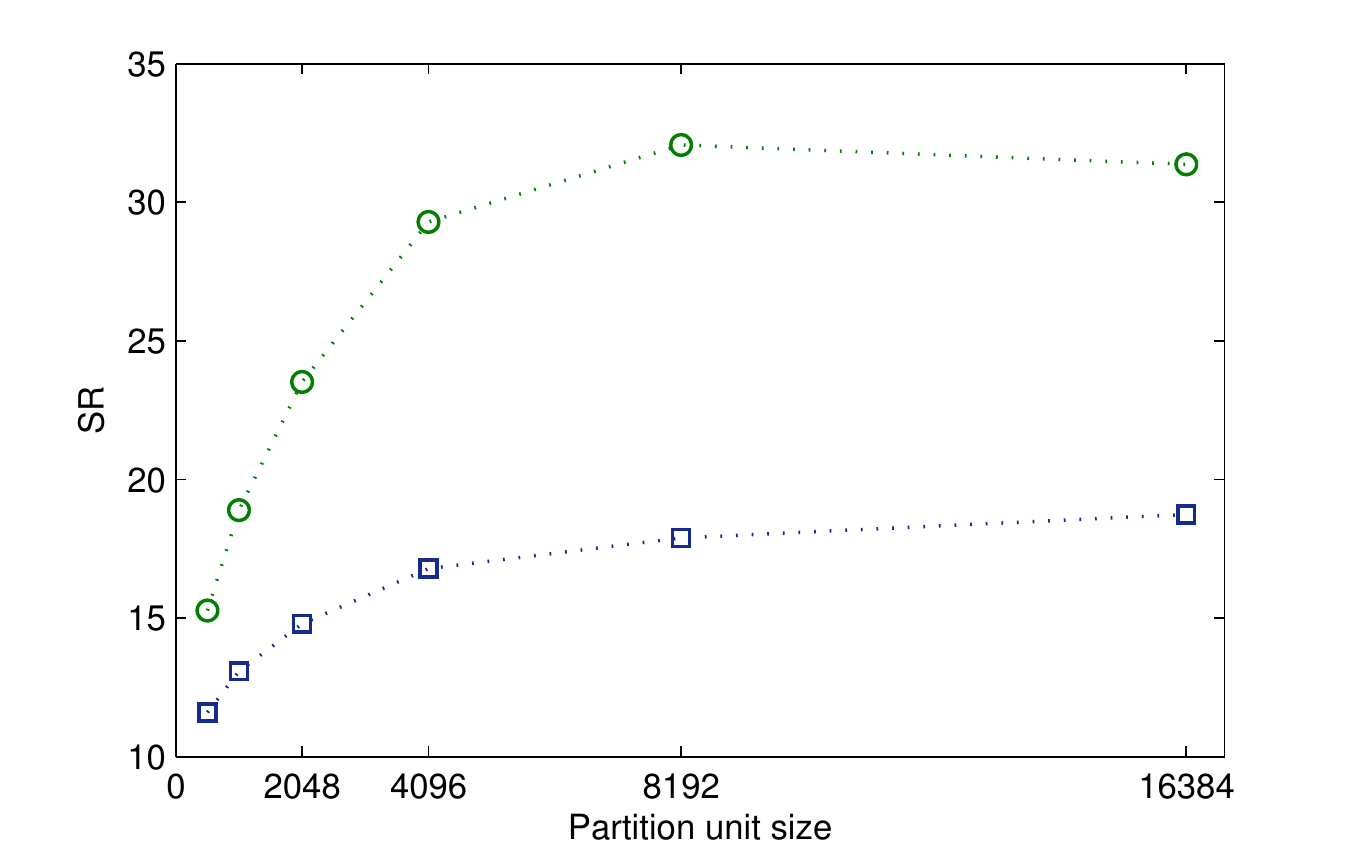}
\caption{\small{(Color online only) SR, for the
Flute Exercise clips (left) and Classical Guitar (right)
corresponding to values
of $\Ns$ equal to 
512, 1024, 2048, 4096, 8192, and 16384 samples.
The squares are the SR values obtained with the orthogonal 
basis $\B^c$. The circles are the results produced by the
mixed dictionary $\D^{cs}$, redundancy four, by means
of the proposed algorithm.}}
\end{center}
\end{figure}
As seen in the figures, 
for all the values of $\Ns$, 
 the gain in sparsity produced by 
the dictionary (represented by the circles in 
Fig.~2) in relation to the best result 
for the basis (squares in those figures) is 
very significant. Table~1  shows the values 
of SR for the clips listed in the 
first column, using the basis 
$\B^c$ and the dictionary $\D^{cs}$ with 
the methods MP and SPMP. The
 value of $\Ns$ is set as that 
producing the best SR for the orthogonal 
basis $\B^c$ which, as illustrated in the left graph of
Fig.~2, is not always the optimal value for the 
dictionary approach. 
 The implementation of the MP algorithm 
via FFT, which we call MPTrgFFT, is ready realized 
simply by deactivating the self projection step. 
\begin{table}[!ht]
\label{tab1}
\begin{center}
\begin{tabular}{|r || r | r| r |r||}
\hline
Clip & $\Ns$ & SR ($\B^c$) & SR (MP) &SR (SPMP)\\
\hline \hline
Flute Exercise&8192&
6.5&
11.8&
13.9\\ \hline
Classic Guitar&16384&
18.7&
26.6&
31.4
\\ \hline
Rock Piano & 2048&
6.9&
10.2&
12.0 \\ \hline
Pop Piano &8192&
11.7 &
15.1 &
18.0
\\\hline
Rock Ballad &8192&
6.8&
8.9&
10.5
\\ \hline
Bach Piano&4096&
11.8&
14.8&
17.4
\\ \hline
Trumpet Solo &8192&
8.3&
11.9&
14.7
\\ \hline
Himno del Riego &4096&
4.9&
7.6&
8.9
\\\hline
Oboe in C &16384&
13.7&
44.1&
53.5
\\\hline
Classical Romance &8192&
7.2  &
11.2 &
13.4
\\\hline
Jazz Organ  &8192&
18.7 &
22.5 &
28.1
\\\hline
Marimba Jazz &1024&
11.8 &
15.3 &
18.6
\\\hline
Begana &2048&
8.5&
10.0&
12.0
\\\hline
Vibraphone &2048&
12.7&
20.1&
23.8
\\ \hline
Polyphon &4096&
3.7 &
6.1 &
7.1
\\\hline
\end{tabular}
\caption{\small{SR 
obtained with the basis $\B^c$ and the 
dictionary $\D^{cs}$, through the MP and SPMP methods, 
for the clips listed in the first column.
The value of the partition unite $\Ns$  
is the one corresponding to the best SR result with the 
basis $\B^c$  when  $\Ns$ takes the 
values 512, 1024, 2048, 4096, 8192, and
 16384.}}
\end{center}
\end{table}
The clips in Table~1 are played with a variety of 
instruments. The sampling frequencies are: 22050 Hz 
 for the Flute Exercise and Himno del Riego,  48000 Hz for the Polyphon, and 44100 Hz for all the other clips. 
 The SR varies significantly, from the 
sparsest clip (Oboe in C) to the least sparse 
one (Polyphon). 
Nevertheless, the gain in sparsity
obtained with the trigonometric dictionaries,
 in relation to the best orthogonal basis,
is in most cases very significant. 
Notice 
that drums are not included in the list. 
The reason being that drum loops are best approximated 
when the  partition size is considerably smaller than  
for the instruments in Table~1. Hence, the 
proposed algorithm is not of particular help in that 
case. On the 
contrary, as discussed in Sec.~\ref{Int}, 
a method linking the approximation of the 
elements in the partition through a global constraint 
on sparsity, or 
quality, is much better suited to that situation 
(Rebollo-Neira 2016a). The same holds true for speech
signals.
Additionally, we understand that drum loops do not fall
 within the class of music that can be sparsely represented
 only with trigonometric atoms of the type we 
are considering here.
 
In order to compare the improvement in 
 SR produced by the SPMP method ($\text{SR}_{\text{SPMP}}$)
 over the MP one ($\text{SR}_{\text{MP}}$)
we defined the relative gain in sparsity as follows: 
\be
\label{gain}
G = \frac{\text{SR}_{\text{SPMP}}- \text{SR}_{\text{MP}}}{\text{SR}_{\text{MP}}} 100 \%
\ee
For the results of Table 1 the mean value gain is 
$\bar{G}= 19.4 \%$  with standard deviation of 2.4\%.
\begin{figure}[!htp]
\begin{center}
\includegraphics[width=8.3cm]{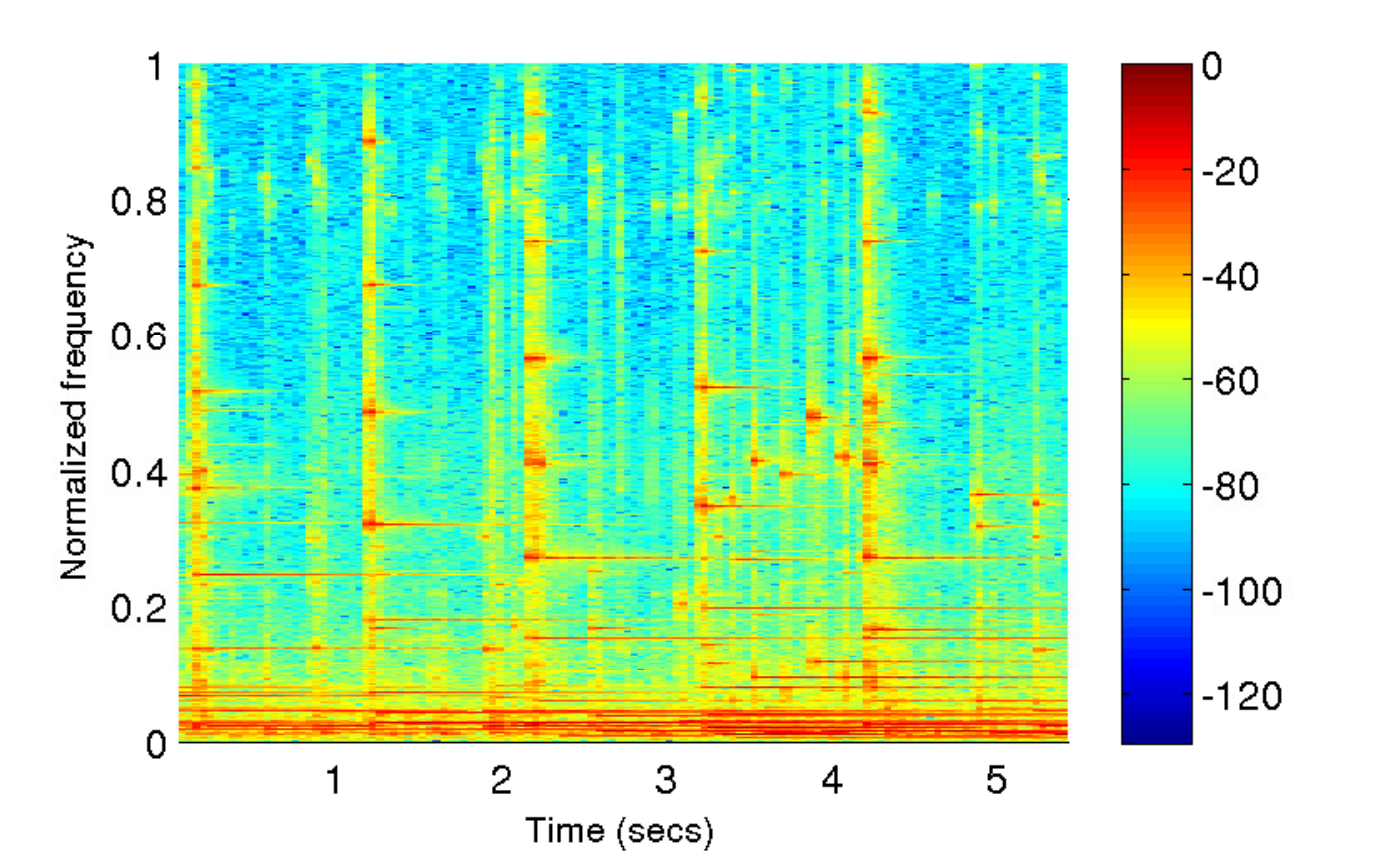}
\includegraphics[width=8.3cm]{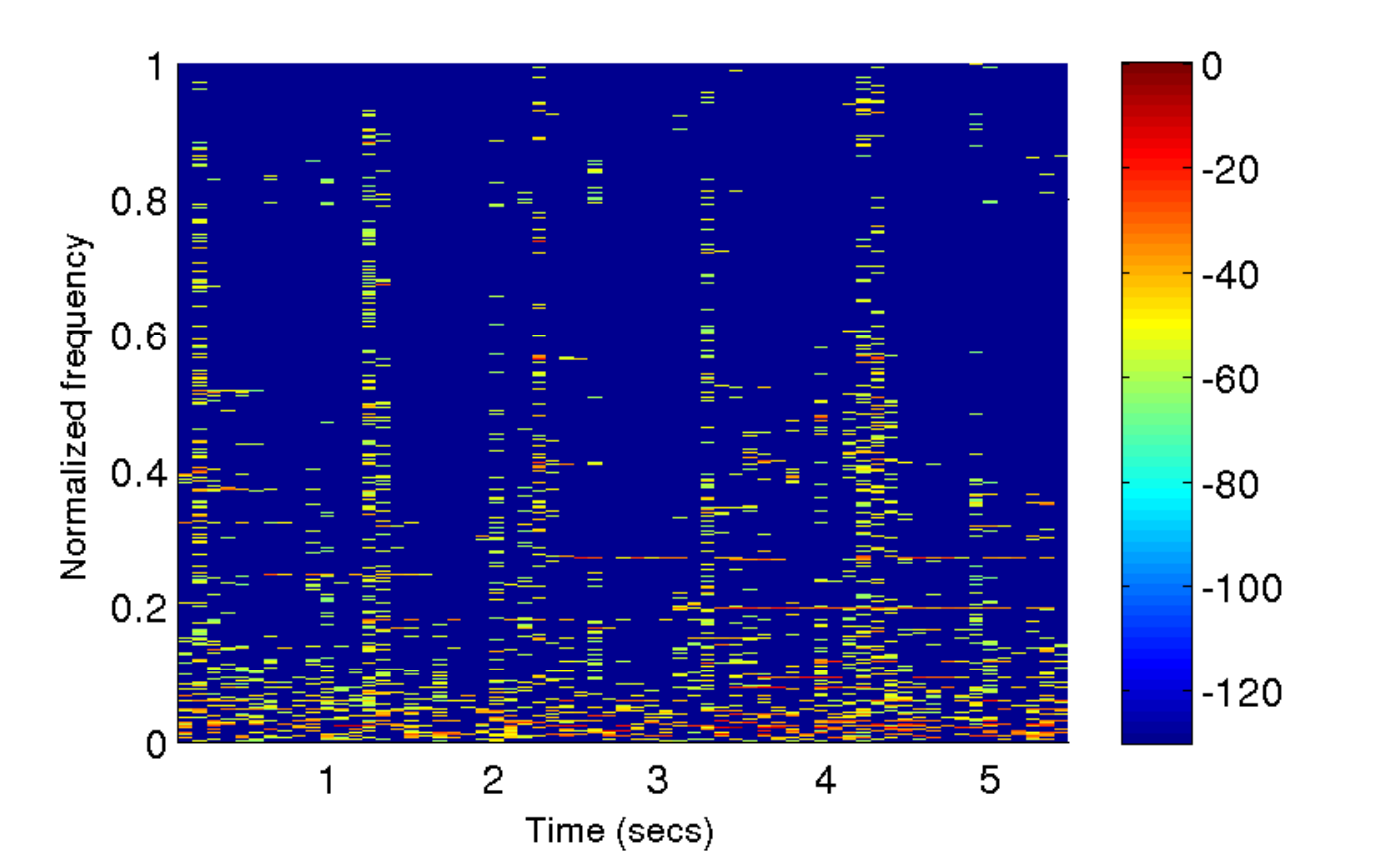}
\caption{\small{(Color online only) 
The left graph is the classic spectrogram of the
Polyphon clip obtained with a Hamming window
of length $4096$ samples and 50$\%$ overlap.
The right graph is the sparser
version of the spectral decomposition,
realized by the trigonometric dictionary and the
 SPMPTrgFFT algorithm,  on a
partition of disjoint units of size $\Ns= 4096$.}}
\end{center}
\end{figure}
Fig.~3 gives a visual representation of the 
 implication of the SR value.
The left graphs  is  a 
classic spectrogram for the Polyphon clip, 
which has been re-scaled to have the maximum value 
equal to one. The right graph is the 
sparse  spectral representation constructed with the 
outputs of the SPMPTrgFFT algorithm (also 
re-scaled to have maximum value equal to one).
Because the spectrograms are given in dB, and the 
sparse one has zero entries, 
the value $10^{-13}$ was added to all the spectral 
power outputs to match scales.

In order to give a description of local sparsity we 
consider the local sparsity ratio 
$sr_q= \frac{\Ns}{k_q},\,q=1,\ldots,Q$, where 
$k_q$ is the number of coefficients in the 
decomposition of the $q$-block and $\Ns$ the size of 
the block. 
 For illustration convenience the graphs in Figs.~4 depict 
  the inverse of this local measure. The  points in 
those figures represent the values 
$1/sr_q,\, q=1,\ldots,Q$.
Each of these values is located in the horizontal axis at 
the center of the corresponding block. 
For each signal the size of the block is taken to be 
the value $\Ns$ yielding the largest SR 
with the dictionary approach, for that particular signal.
\begin{figure}[!htp]
\begin{center}
\includegraphics[width=8.3cm]{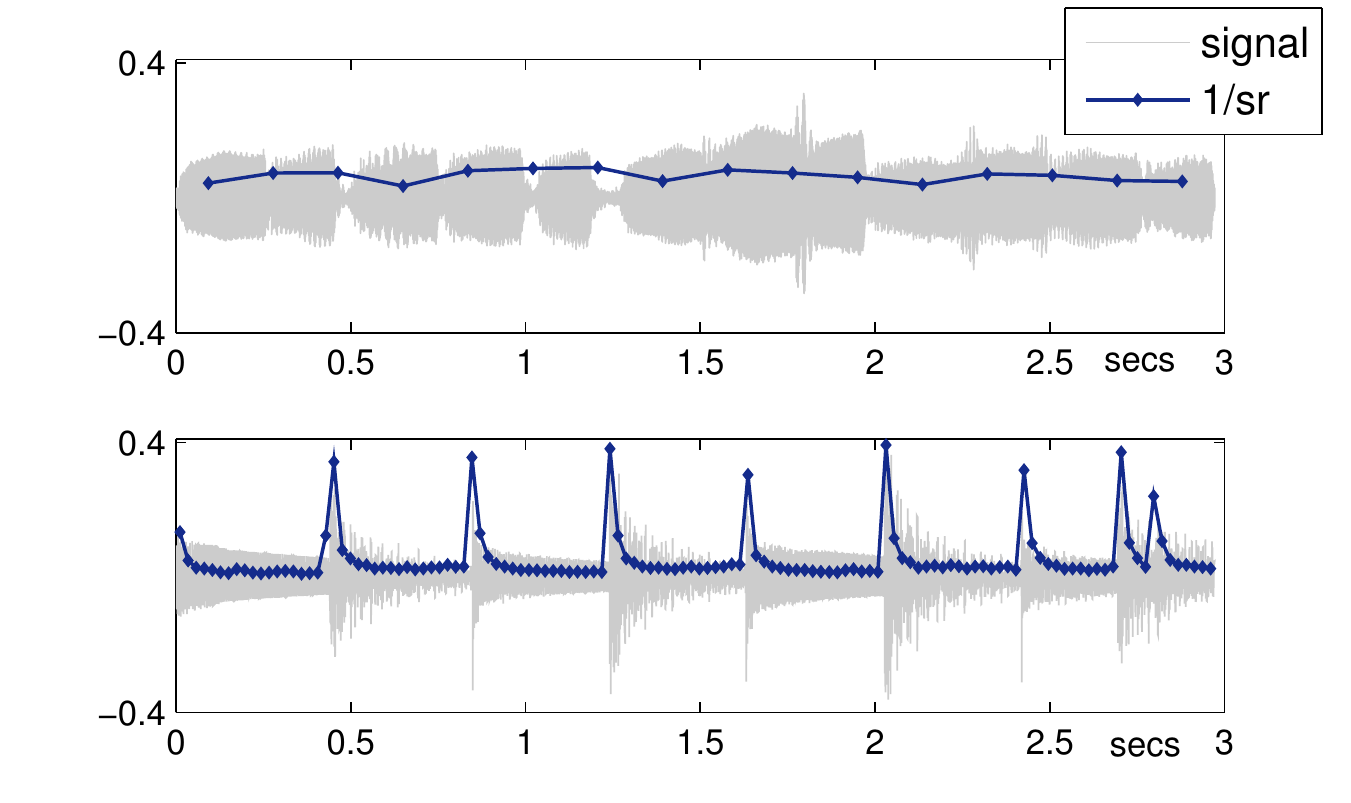}
\includegraphics[width=8.3cm]{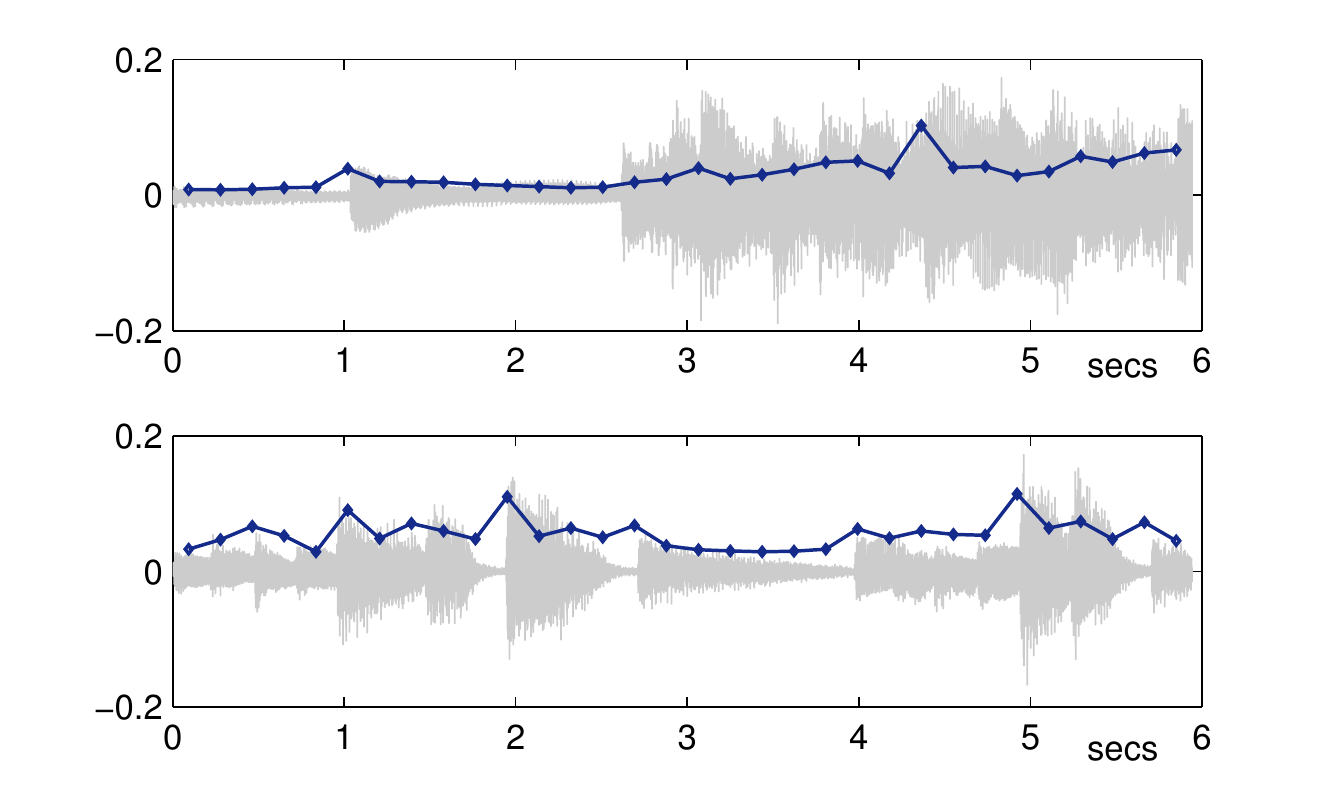}
\caption{\small{(Color online only) The points joined by the darker line in
all the graphs are the
values of the inverse local sparsity ratio
 $1/sr_q,\,q=1,\ldots,Q$. The top graphs correspond to
the Flute (left) and Classic Guitar clips.
The bottom graphs correspond to the Marimba (left) and
Pop Piano clips. The lighter lines represent the
signals.}}
\end{center}
\end{figure}

 The lighter lines in all the graphs of Fig.~4 
 represent the Flute, Marimba   
Classic Guitar and Pop Piano clips. It is interesting to 
see that each if the darker lines joining the inverse local 
sparsity points  follows,  somewhat,
 the shape of signal's envelop.  
This is particularly noticeable when a transient occurs. 

As opposed to the method of Serra and Smith (1990),
which would model a possible component of a sound clip
by tracking the evolution of some
frequencies along time, but in general would produce
a significant residue, the goal of the proposed 
sparse spectral representation is to achieve high
quality reconstruction. As indicated by the points in 
the graphs of Fig.~4, for some signals this is attained 
by a decomposition of low local sparsity in particular 
blocks.  
Notice, however, that 
a signal exhibiting such picks of inverse 
local sparsity  may 
produce, on the whole, a SR which is higher than 
 the SR of a signal endowed with more uniform 
local sparsity, e.g. Flute vs Marimba and Pop Piano.
%
%
The clips of Table~1 are all played with single instruments. 
The rather high value of SNR (35dB) is 
set to avoid noticeable loss or artifacts in the 
signal reconstruction, which might be easy to 
detect due to the nature of the sound. 
 Nevertheless, for  the 
clips of Table~2, which are played by multiple instruments,  
 for  SNR$=$25dB (and even lower) we do not perceive 
loss or artifacts. Hence, the sparsity results 
of Table~2 correspond to SNR=25dB. Overestimating the 
 required SNR for high quality recovery would produce a 
  significant reduction of the SR values.
\begin{table}[!htp]
\label{tab2}
\begin{center}
\begin{tabular}{|r || r| r |r||}
\hline
Clip & SR ($\B^c$) & SR (MP) &SR (SPMP)\\
\hline \hline
Classic Music (sextet)&
12.2&
16.2&
18.4\\ \hline
Piazzola Tango (quartet)&
10.7&
13.8&
15.7
\\ \hline
Opera (female voice)&
5.6&
7.5&
8.3
\\ \hline
Opera (male voice)&
9.2&
12.0&
13.5
\\ \hline
Bach Fugue (orchestral version)&
8.2&
12.4&
14.1
\\ \hline
Simple Orchestra&
13.1&
17.6&
19.8
\\ \hline
\end{tabular}
\caption{\small{SR 
obtained with the basis $\B^c$ and the
dictionary $\D^{cs}$, through the MP and SPMP methods,
for the clips listed in the first column.
The partition unite size is in all the 
cases $\Ns=4096$ and the sampling frequency  
44100 Hz.}}
\end{center}
\end{table}\\
For the results of Table 2  the mean value gain 
in SR (c.f. \eqref{gain}) is $\bar{G}= 12.8 \%$  with 
standard deviation of $1.2\%$.

{\bf{Remarks on computational complexity:}}
The increment in the computational complexity  of SPMPTrgFFT 
with respect to MPTrgFFT
 is a 
factor  which accounts for the iterations 
realizing the self-projections.
In order to 
estimate the complexity we
indicate by $\ov{\ov{\kappa}}$ the 
double average 
of the number of iterations in the projection step.
More specifically, indicating by $\kappa_k$ the 
number of iterations in the $k$-term approximation 
of a  fixed segment $q$, $\ov{\kappa}_q = \frac{1}{k_q}
\sum_{k=1}^{k_q} \kappa_k$  and
${\ds{\ov{\ov{\kappa}}= \frac{1}{Q} 
\sum_{q=1}^Q \ov{\kappa}_q}}$. 

The value of $\ov{\ov{\kappa}}$ gives an estimation of the 
SPMPTrgFFT complexity: 
O($\ov{\ov{\kappa}} K M \log_2 M$). 
Since for a dictionary of 
redundancy $r$  the number of elements is $M=r \Ns$, 
in order to make clearer the influence of 
the segment's length in the complexity, 
this can be expressed as
O($\ov{\ov{\kappa}} K r \Ns \log_2 r \Ns)$.
The computational complexity of plain MPTrgFFT 
is given by the complexity of calculating 
 inner products via FFT, i.e. 
 O($ K r \Ns \log_2 r \Ns)$. Hence 
 $\ov{\ov{\kappa}}$ 
gives a measure  of the increment of complexity 
introduced by the projections to achieve the desired 
optimality in the 
coefficients of the approximation.
Fig.~5 shows the values of $\ov{\ov{\kappa}}$ 
as a function of the segment's length $\Ns$.
The triangles correspond to the Flute Exercise clip  
the starts to the Classic Guitar clip.  
Notice that for the Flute Exercise 
the value of $\ov{\ov{\kappa}}$ 
augments significantly for the two larger values of $\Ns$,
while remains practically constant 
for the Classic Guitar. This feature is in line 
with the fact that, as seen in 
Fig~2, the SR for those values of $\Ns$
is practically constant for the Classic Guitar, 
but decreases for the Flute Exercise. 

\begin{figure}[!ht]
\begin{center}
\includegraphics[width=9cm]{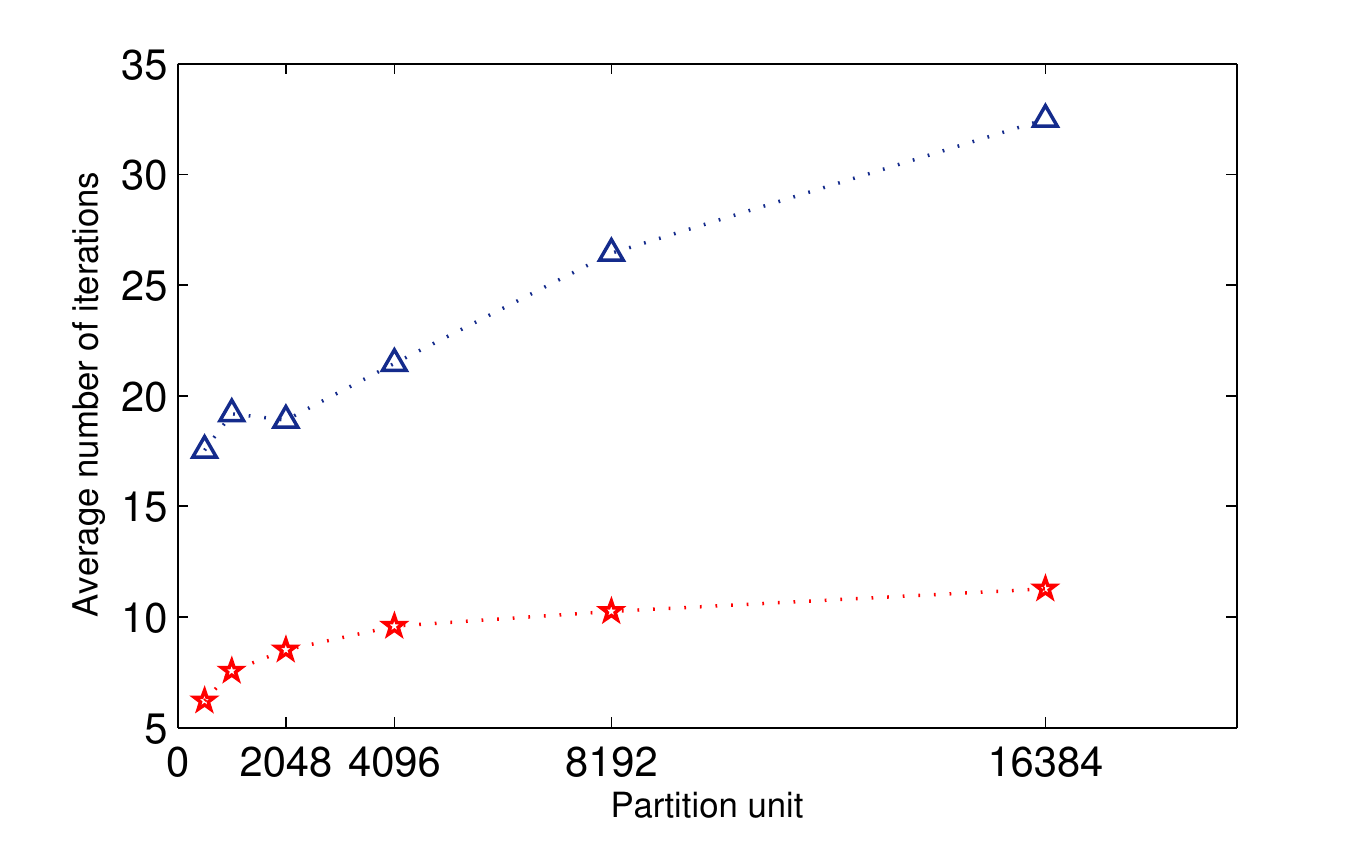}
\caption{\small{(Color online only) Average number of the iterations, 
$\ov{\ov{\kappa}}$, for 
realizing the projection step procedure 
 (Algorithm  5) 
corresponding to partition units of length  
 $\Ns$ equal to 512, 1024, 2048, 4096, 8192, 
and 16384 samples. 
The triangles are the values for the flute clip  and 
the starts for the classic guitar.
}}
\end{center}
\end{figure}
\section{Conclusions}
\label{Con}
A dedicated method for sparse spectral representation of
music sound 
has been presented. The method was devised for the 
representation to be realized outside the orthogonal basis 
framework. Instead, the spectral components are selected 
from an overcomplete trigonometric dictionary. 
The suitability of these dictionaries for sparse 
representation of melodic music, by partitioning, 
 was illustrated on a 
number of sound clips of different nature. 
While the quality of the reconstruction is an 
input of the algorithm, the method is conceived 
to achieve high quality recovery. Hence, 
in order to benefit sparsity results the signal
 partition is realized without overlap. 
The approach has been shown to be worth  
applying to improve sparsity within the class of
signal which are compressible in terms of a 
trigonometric basis. 
The achieved sparsity 
is theoretically equivalent to that produced by the 
OMP approach with the identical dictionary. The numerical 
equivalence of both algorithms was verified when possible. 

In order to facilitate the application of the approach 
 we have made publicly available the MATLAB version of 
Algorithms 1-6  on a dedicated web page$^1$.
It is appropriate to stress, though, that the 
routines are not intended to be an optimized
implementation of the method. On the contrary, 
they have been produced with the intention of 
providing an easy to test form of the 
approach. We hope that the MATLAB version of the  
algorithms will facilitate  
their implementation in appropriate 
programming languages for practical applications.

\subsection*{Acknowledgements}
We are grateful to three anonymous reviewers for 
many comments and suggestions for improvements to previous 
 versions of the manuscript. We are also 
grateful to Xavier Serra who has kindly let us have  
a MATLAB function for the implementation of their 
 method (Serra and Smith, 1990).  
\theendnotes{}
\newpage
\appendix
\subsection*{Appendix A}
\newcounter{myalg}
\begin{algorithm}[!htp]
\refstepcounter{myalg}
\begin{algorithmic}
\label{IP}
\caption{Computation of inner product with  a
trigonometric dictionary via FFT.
IPTrgFFT procedure: $[\vIP]=$IPTrgFFT($\vRe, M,$ Case)}
\STATE{\bf{Input:}}$\,\vR \in \R^N$, $M$, number of
elements in the dictionary, and Case (I , II, or III).\\

\STATE \COMMENT{{Computation of the inner products
$\vIP = \la \vd , \vRe \ra \in \C^M$}}

\STATE{Case I}

\STATE $\vIP={\rm{FFT}}(\vR, M$)$\frac{1}{\sqrt{N}}$,

\STATE Case II, III (c.f. \eqref{dct}, \eqref{dst})

\STATE \COMMENT{Computation of auxiliary vector
$\Aux \in \C^{2M}$ to compute $\vIP$.}

\STATE{$\Aux={\rm{FFT}}(\vRe, 2 M)$}

\STATE{Case II}
\STATE{$IP(n)=\frac{1}{w^c(n)} 
\Re(\expn Aux(n)),\, n=1,\ldots,M$}

\STATE{Case III}

\STATE{$IP(n-1)=-\frac{1}{w^s(n)}\Im(\expn Aux(n)),\, n=2,\ldots,M+1$}
\label{AlIP}
\end{algorithmic}
\end{algorithm}
\begin{algorithm}[!htp]
\refstepcounter{myalg}
\label{TrgAt}
\begin{algorithmic}
\caption{Generation of an atom, given the
index and the dictionary type.
Trigonometric Atom procedure:
$[\vd_{\ell_k}]$=TrgAt($\ell_k, M, N,$ Case)}

\STATE{\bf{Input}}:\, Index $\ell_k$, number of
elements in the dictionary $M$, atom's dimension $N$,
Case (I, II, III or IV).

\STATE{{\bf{Output}}:\, Atom $\vd_{\ell_k}$.}

\STATE \COMMENT{Generation of the atom, $\vd_{\ell_k}$,
according to the Case}

\IF {Case=IV}

\STATE{$M \leftarrow \frac{M}{2}$}

\ENDIF

Case I\\

${\ds{d_{\ell_{k}}(j)=\frac{1}{\sqrt{N}}e^{\imath \frac{{2\pi(j-1)(\ell_k-1)}}{M}},\quad
j=1,\ldots,N}}$

Case II (and Case IV if $\ell_{k} \leq \frac{M}{2}$)

${\ds{d_{\ell_{k}}(j) = \frac{1}{w^c(\ell_k)} \,
 \cos ({\frac{{\pi(2j-1)(\ell_k-1)}}{2M}}), \quad
j=1,\ldots,N}}$\\

Case III (and Case IV if ${\ds{\ell_{k} > \frac{M}{2}}}$)

${\ds{d_{\ell_{k}}(j)= \frac{1}{w^s(\ell_k)} \,
 \sin ({\frac{{\pi(2j-1)\ell_k}}{2M}}), \quad
j=1,\ldots,N}}$
\label{TrgA}
\end{algorithmic}
\end{algorithm}
\begin{algorithm}[!htp]
\label{AtSel}
\refstepcounter{myalg}
\caption{{Atom Selection via FFT.
AtSelFFT procedure:
$[\ell_k, c(\ell_k)]=$AtSelFFT($\vR,M,$ Case)}}
\begin{algorithmic}
\STATE{\bf{Input:}}\, Residual $\vR \in \R^{N}$,
$M$ number of elements in the dictionary,
and Case (I, II, III, or IV)

\STATE{\bf{Output:}}\,Index of the selected atom $\ell_k$,
 and  MP coefficient
$c(\ell_k)= \la \vd_{\ell_k}, \vR \ra$
calculated via FFT.

\COMMENT{Call IPTrgFFT procedure, Algorithm \ref{AlIP},
to calculate inner products}

\STATE{Case I}

$\vIP$=IPTrgFFT($\vRe,M,$ Case I),

\STATE{Cases II and III}

$\vIP$=IPTrgFFT($\vRe, M,$ Case),

\COMMENT{Selection of the new atom and evaluation of the MP
coefficient}

${\ds{\ell_{k}= \operatorname*{arg\,max}_{\substack{n=1,\ldots,M}}|IP(n)|}}$

$c(\ell_{k})=IP(\ell_{k})$\\

\STATE{Case IV}

$M \leftarrow \frac{M}{2}$

\STATE{$\vIP^c$=IPTrgFFT($\vRe,M$, Case II)\\

$\vIP^s$=IPTrgFFT($\vRe,M,$ Case III)}

$\ds{\nu= \operatorname*{max}( |IP^c(\ell^c)|, 
 |IP^s(\ell^s)|),\,\,
\text{with}\,\,
\ell^c= \operatorname*{arg\,max}_{\substack{n=1,\ldots,M}}|IP^c(n)|\,\, 
\text{and\,\,}
\ell^s= \operatorname*{arg\,max}_{\substack{n=1,\ldots,M}}|IP^s(n)|}$
\IF {$\nu= |IP^s(\ell^s)|$}
\STATE{$\ell_{k}=\ell^s + M $ and $c(\ell_{k})= IP^s(\ell^s)$}
\ELSE{
\STATE{$\ell_{k}=\ell^c$ and $c(\ell_{k})= IP^c(\ell^c)$}}
\ENDIF
\end{algorithmic}
\end{algorithm}
\begin{algorithm}[!htp]
\refstepcounter{myalg}
\caption{{Atom Re-Selection via FFT.
AtReSelFFT procedure:
$[\ell, c(\ell)]$=AtReSelFFT($\vR,M,\Gamma,$ Case)}}
\label{AlReSel}
\begin{algorithmic}
\STATE{\bf{Input:}}\, Residue $\vR \in \R^{N}$,
 number of dictionary's
elements, $M$, set of indices of the selected atoms  $\Gamma=\{\ell_n\}_{n=1}^{k}$ (if Case=IV both, 
$\Gamma^c$,
 indices for atoms in $\D^c$, and $\Gamma^s$, 
 indices for atoms in $\D^s$).

\STATE{\bf{Output:}}$\,$ Re-Selected index $\ell$
(out of the set $\Gamma$) and corresponding
MP coefficient $c(\ell)=\la \vd_{\ell}, \vR\ra, \ell \in \Gamma$, calculated via FFT.


\STATE{Case I}

$\vIP$=IPTrgFFT($\vRe, M,$ Case I),

\STATE{Cases II and III}

$\vIP$=IPTrgFFT($\vRe, M,$ Case ),

\COMMENT{Selection of the index $\ell \in \Gamma$}

${\ds{\ell= \operatorname*{arg\,max}_{\substack{n \in \Gamma}}|IP(n)|}}$

$c(\ell)=IP(\ell)$
\STATE{Case IV}

$M \leftarrow \frac{M}{2}$

\STATE{$\vIP^c$=IPTrgFFT($\vRe,M,$ Case II)\\

$\vIP^s$=IPTrgFFT($\vRe,M,$ Case III)}

$\ds{\nu= \operatorname*{max}(|IP^c(\ell^c)|,|IP^s(\ell^s)|,\,\,
\text{with}\,\,
\ell^c= \operatorname*{arg\,max}_{\substack{n \in \Gamma^c}}|IP^c(n)|\,\, 
\text{and\,\,}
\ell^s= \operatorname*{arg\,max}_{\substack{n \in \Gamma^s}}|IP^s(n)|}$

\IF {$\nu= |IP^s(\ell^s)|$}

\STATE{$\ell=\ell^s + M$ and $c(\ell)= IP^s(\ell^s)$}

\ELSE{

\STATE{$\ell=\ell^c$ and $c(\ell)= IP^c(\ell^c)$}}

\ENDIF

\end{algorithmic}
\end{algorithm}

\begin{algorithm}[!htp]
\refstepcounter{myalg}
\begin{algorithmic}
\caption{Orthogonal Projection via FFT.
ProjMPTrgFFT procedure:
$[\vRt,\vct]$=ProjMPTrgFFT($\vRe, M, \vc, \Gamma,\epsilon,$ Case)}
\label{ProjMP}
\STATE{\bf{Input:}}\, Residue
$\vRe \in \R^{N}$,
number of elements in the
dictionary, $M$, vectors $\vc$ with
the coefficients in the $k$-term approximation,
set  $\Gamma $ of selected indices up to iteration $k$, tolerance for the
numerical error of
the projection $\epsilon$, and Case (I, II, III, or IV).

\STATE{\bf{Output:}}\, Updated residue,
$\vRt \in \R^{N}$, orthogonal to
$\Spann\{\vd_n\}_{n \in \Gamma}$ and updated coefficients
$\vct$ accounting for the projection.

\COMMENT{Set $\mu= 2 \epsilon$ to start the algorithm}

\WHILE {$\mu > \epsilon$}
\STATE\COMMENT{Select one index from $\Gamma$ to construct
the approximation of $\vRe$ in $\Spann\{\vd_n\}_{n \in \Gamma}$}

\STATE{$[\ell,\ct(\ell)]$=AtReSelFFT($\vRe, M, 
 \Gamma,$ Case)}

\COMMENT{Generate the selected atom $\vd_{\ell}$}

\STATE{$\vd_{\ell}$=TrgAt($\ell, M, N$, Case).}

\COMMENT{Update residue}

\STATE{${\ds{\vR \leftarrow\vR - \ct(\ell) \vd_\ell}}$}

%

\COMMENT{Since $\vR$ is vector of real numbers}

    \IF {Case = I}

\STATE{ $\ell'=M- \ell+2$,

      $\vd_{\ell'}$=TrgAt($\ell', M, N$, Case),

      ${\ds{\vR} \leftarrow \vR - \ct^\ast(\ell)\vd_{\ell'}}$
}
    \ENDIF

\STATE{$\mu=|\ct(\ell)|$}

\COMMENT{Update coefficient}

$c(\ell) \leftarrow c(\ell) + \ct(\ell)$

\IF {Case = I}

\STATE{$c(M -\ell +2) \leftarrow c^\ast(\ell)$}

\ENDIF

\ENDWHILE

\COMMENT{Rename coefficients and residue to match the
output variables}

\STATE{$\vct=\vc,\, \vRt=\vR$}
\end{algorithmic}
\end{algorithm}
\begin{algorithm}[!htp]
\refstepcounter{myalg}
\begin{algorithmic}
\caption{Main Algorithm for the proposed
SPMP method dedicated to trigonometric dictionaries and
implemented via FFT.
Procedure SPMPTrgFFT: $[\vf^k, \vc, \Gamma]=$
SPMPTrgFFT($\vf, M, \rho, \epsilon,$ Case)}
\label{algo}
\STATE{\bf{Input:\,}} Data $\vf \in \R^{N}$, $M$, number of
elements in the dictionary,
approximation error $\rho>0$ and tolerance $\epsilon>0$
for the numerical realization of the projection
 Case (I , II, III, or IV).

\STATE{\bf{Output:$\,$}}
Approximated data $\vf^k \in  \R^{N}$.
Coefficients in the atomic decomposition, $\vc$,
Indices labeling the selected atoms
$\Gamma=\{\ell_n\}_{n=1}^k$.\\

\STATE \COMMENT{{Initialization}}

\STATE{\rm{Set}}$\,\,\Gamma\,=\,\{\emptyset\},\quad \vf^0\,=\,0,\quad
\vR^0 \,=\,\vf,\,\quad k\,=\,0, \quad \mu= 2\rho$


\STATE\COMMENT{{Begin the algorithm}}

\WHILE {$\mu > \rho$}

\STATE $k=k+1$

\COMMENT{Select index
$\ell_k$ and  calculate  $c(\ell_k)$}

\STATE$[\ell_k, c(\ell_k)]$=AtSelFFT($\vRe^{k-1}, M,$ Case)

\COMMENT{Generate the atom $\d({\ell_k})$}

$\vd_{\ell_k}$=TrgAt$(\ell_k, M, N,$ Case)

\STATE Updated $\Gamma \leftarrow \Gamma \cup \ell_k$

\COMMENT{Calculate approximation and residue}

\STATE $\vf^k = \vf^{k-1} + c(\ell_k) \vd_{\ell_k},$
\text{and} $\vR^k  = \vf - \vf^k$

\COMMENT{Subtract from $\vR^k$ the component
in $\Spann\{ \vd_n\}_{n \in \Gamma}$}

[$\vRt^k,\vct$]=ProjMPTrgFFT($\vRe^{k}, 
M, \vc, \Gamma,\epsilon,$ Case)

\COMMENT{Update residue, approximation, coefficients, and error}

\STATE $\vR^{k} = \vRt^k, \vf^k =\vf - \vR^k; \vc = \vct, \mu  = \|\vR^{k}\|$
\ENDWHILE
\end{algorithmic}
\end{algorithm}
\newpage

\newpage
\listoffigures
\end{document}